\documentclass[conference]{IEEEtran}
\IEEEoverridecommandlockouts
\usepackage{cite}
 \usepackage[utf8]{inputenc}
     \usepackage{setspace}
\usepackage{amsmath,amssymb,amsfonts}
\usepackage{dirtytalk}
\usepackage{algorithmic}
\usepackage{graphicx}
\usepackage{textcomp}
\usepackage{xcolor}

\usepackage{gensymb}

\usepackage[flushleft]{threeparttable}
\usepackage{makecell,booktabs}

\usepackage[english]{babel} 
\usepackage{blindtext}
\newcommand{\RNum}[1]{\uppercase\expandafter{\romannumeral #1\relax}}

\def\BibTeX{{\rm B\kern-.05em{\sc i\kern-.025em b}\kern-.08em
    T\kern-.1667em\lower.7ex\hbox{E}\kern-.125emX}}
\begin{document}

\title{Terahertz Communication Testbeds: Challenges and Opportunities
}

\author{
}

\author{\IEEEauthorblockN{Eray Güven, Güneş Karabulut Kurt} 
\IEEEauthorblockA{Poly-Grames Research Center, Department of Electrical Engineering\\ Polytechnique Montr\'eal, Montr\'eal, Canada \\
E-mail: guven.eray@polymtl.ca, gunes.kurt@polymtl.ca
}
}

\maketitle

\begin{abstract}
This study investigates an experimental software defined radio (SDR) implementation on $180$ $\mathrm{GHz}$. \textcolor{black}{The system model is presented to evaluate the performance and unveil hidden opportunities. Accordingly, rate scarcity and frequency sparsity are discussed as hardware bottlenecks.} Multiple error metrics for the terahertz (THz) signal are acquired, and various case scenarios are subsequently compared, revealing that the SDR-THz testbed reaches $3.2$ $\mathrm{Mbps}$ with $<1^{\circ}$ skew error. \textcolor{black}{It is observed that} the use of a reflector plate can fine-tune the frequency error and gain imbalance in the expense of at least $14.91$ $\mathrm{dB}$ signal-to-noise ratio. The results demonstrate the complete feasibility of SDR-based baseband signal generation in THz communication, revealing abundant opportunities to overcome hardware limitations in experimental research.
\end{abstract}


\section{Introduction}
\label{intro}
Advancements in broadband communication open up numerous opportunities in wireless communication, robotics, sensing, spectroscopy and astronomy. One of the pivotal enablers, terahertz (THz) communication, offers the potential for communication rates in the terabits per second ($\mathrm{Tbps}$) range, making it a viable long-term solution for alleviating congestion \cite{jiang2023terahertz}. Analytical studies reveal that aerial networks \cite{yuan2022secure}, wireless backhaul \cite{sen2023multi}, and indoor communication \cite{shafie2020multi} are increasingly gravitating towards high-throughput and secure utilization of THz technology.
Nonetheless, it has been observed that the subject remains relatively limited in terms of experimental research, primarily because of hardware limitations \cite{sarieddeen2021overview}.

The frequency range of $1-10$ $\mathrm{THz}$ is commonly referred to as the ``THz region", whereas the carrier frequency ($f_c$) range of $0.1-1$ $\mathrm{THz}$ is often termed the ``sub-THz" region. In contrast to THz communication, sub-THz communication combines the characteristics of mm-Wave communication with the propagation features of THz waves. As a result, it occupies a distinctive niche within wireless communication, finding applications ranging from terrestrial links \cite{xing2021terahertz} to satellite communications \cite{kumar2022snr}. Exposing the line-of-sight (LoS) environment, sub-THz communication can be even implemented over recently proposed inter-satellite communication links\cite{guven2023multi}. Regarding the experimental studies, sub-THz testbeds have been conducted recently in \cite{sen2020teranova} and \cite{Shehata2022}. 
It can be seen that these studies are significantly application specific and require heavy digital signal processing (DSP) capability. We underscore the necessity for a testbed that combines versatility with the inherent characteristics of THz communication. \textcolor{black}{In \cite{guven2024mutuality}, the directive nature of sub-THz communication is empirically leveraged for localization purposes.  }

A software-defined radio (SDR) is a versatile radio frequency (RF) equipment with the capability to process signals during both transmission and reception. It facilitates real-time experimentation with intricate computations, encompassing tasks such as phase shifting, rate splitting, as well as source and channel coding. Thence, beam management and multi-access, spectrum surveillance, are some of the concrete use cases of SDRs on the physical layer. The recent developments on radio-frequency integrated circuit (RFIC) and programmable logic units show that SDR wireless testbeds are capable of conducting high complexity designs such as beamforming \cite{toka2021performance}, space-time block coding \cite{zhang2019design} and even broadband communication \cite{ariyarathna2023toward}. The study \cite{ariyarathna2023toward} presents a system-on-chip based DSP solution to the rate scarcity problem for ultra-broadband sub-THz SDR tests, which allows multi-band transmission. However, the utilization of frequency multiplexing sampling for multi-band channelization results in intra-band gaps, leading to a reduction in spectral efficiency. Accordingly, the hardware bottlenecks emerge as a novel investigation for SDR based THz testbeds.

SDR technology presents a wide range of possibilities for real-time THz prototypes. Nevertheless, the high demand requirements for THz communication cannot be met by the current SDR technology, such as ultra-broadband transmission and very high data streaming with sampling rate ($f_s$) of $\mathrm{Tbps}$ level. Nonetheless, besides providing design flexibility, SDR-assisted sub-THz and THz technologies enable channel and wave propagation modeling with digital signal processing techniques. However, an SDR for baseband signal generation with $\sim 200$ $\mathrm{Mbps}$, replacing an arbitrary wave generator (AWG) of $>200$ $\mathrm{Gbps}$, is a major challenge. The host interface capacity and signal processing over ultra-broadband signal are two main hardware constraints. \textcolor{black}{Once these limitations are relaxed, the SDR-THz testbed can practically assess the performance of real-time multi-gigabit connectivity applications, such as virtual and augmented reality, ultra-secure connectivity, and high-capacity uses like wireless backhaul.}

In order to mitigate the aliasing for Nyquist criteria, oversampling and anti-aliasing baseband filters are two common solutions to implement in the real-time test environment. Sampling in Nyquist frequency range via an SDR, intermediate frequency (IF) signal can be up-converted with an either exterior heterodyne mixer or a harmonic multiplier. This model also requires a down conversion unless the receiver has the capability with near $\mathrm{Tbps}$ rate analog-to-digital converter. Today's analog-to-digital converter (ADC) technology is beyond $200$ $\mathrm{Gbps}$, further ahead of broadband filtering and RFIC technology, but not as cost-efficient. Thankfully, all broadband-related challenges, such as buffering, can be surmounted by converting the THz carrier back to IF.   

In order to enhance the efficient utilization of THz propagation and mitigate the effects of environment exposed THz, one approach is the implementation of reconfigurable intelligent surfaces (RIS). RIS are state-of-the-art programmable structures that can be utilized in both mm-Wave and THz \cite{le2023performance}. Apart from reflection, it has potential use with beamforming, beamshaping and amplifying the incoming wave. Similarly, reflectors are easy to fabricate structures that are commonly used in experimental wireless designs for several purposes. Essentially, reflectors are able to evaluate the scattering and non-line of sight (NLoS) communication performance. Interestingly, reflectors also can imitate the passive RIS without phase shifter elements. A sub-THz study \cite{jornet2023wireless} highlights the impact of a reflector on the wavefront experimentally. While \cite{jornet2023wireless} explores the impact of reflections use on THz experiments, there is still a gap in the literature regarding the development of a solid understanding of the use of SDRs in the THz band.

The main contributions of this study can be summarized as following:
\begin{itemize}
    \item A realistic system model with non-ideal transmitter model for THz is derived. Correspondingly, an SDR based THz experiment is conducted. Key challenges and hardware constraints of such a cascaded platform are discussed.
    \item Hardware bottlenecks of SDR-THz testbed have been unveiled. Rate scarcity and frequency sparsity for SDR-THz applications are expanded. The stability requirements for ultra-broadband end-to-end communication are addressed.
    \item The utilization of reflectors in an SDR-THz testbed is examined. A distance-based relationship with the reflectors is established and the benefits they offer are highlighted.
    \item In-phase and quadrature (I/Q) level error metrics are obtained such as signal-to-noise ratio (SNR), error vector magnitude (EVM), gain imbalance, amplitude droop, skew and frequency error. 
\end{itemize}

The rest of the paper is organized as follows: Section \ref{sec2} explains the proposed generic SDR-THz system model with hardware noises. In Section \ref{sec3}, two major hardware bottlenecks for SDR-THz are discussed. Section \ref{exp} presents the experiment testbed instrumentation and performance evaluation. \textcolor{black}{Lastly, we discuss the open issues within the experimental THz studies and conclude the paper in Section \ref{sec6}.}
\section{SDR-THz Testbed}
\label{sec2}
\begin{figure}[] 
    \centering
    \includegraphics[width=1\columnwidth]{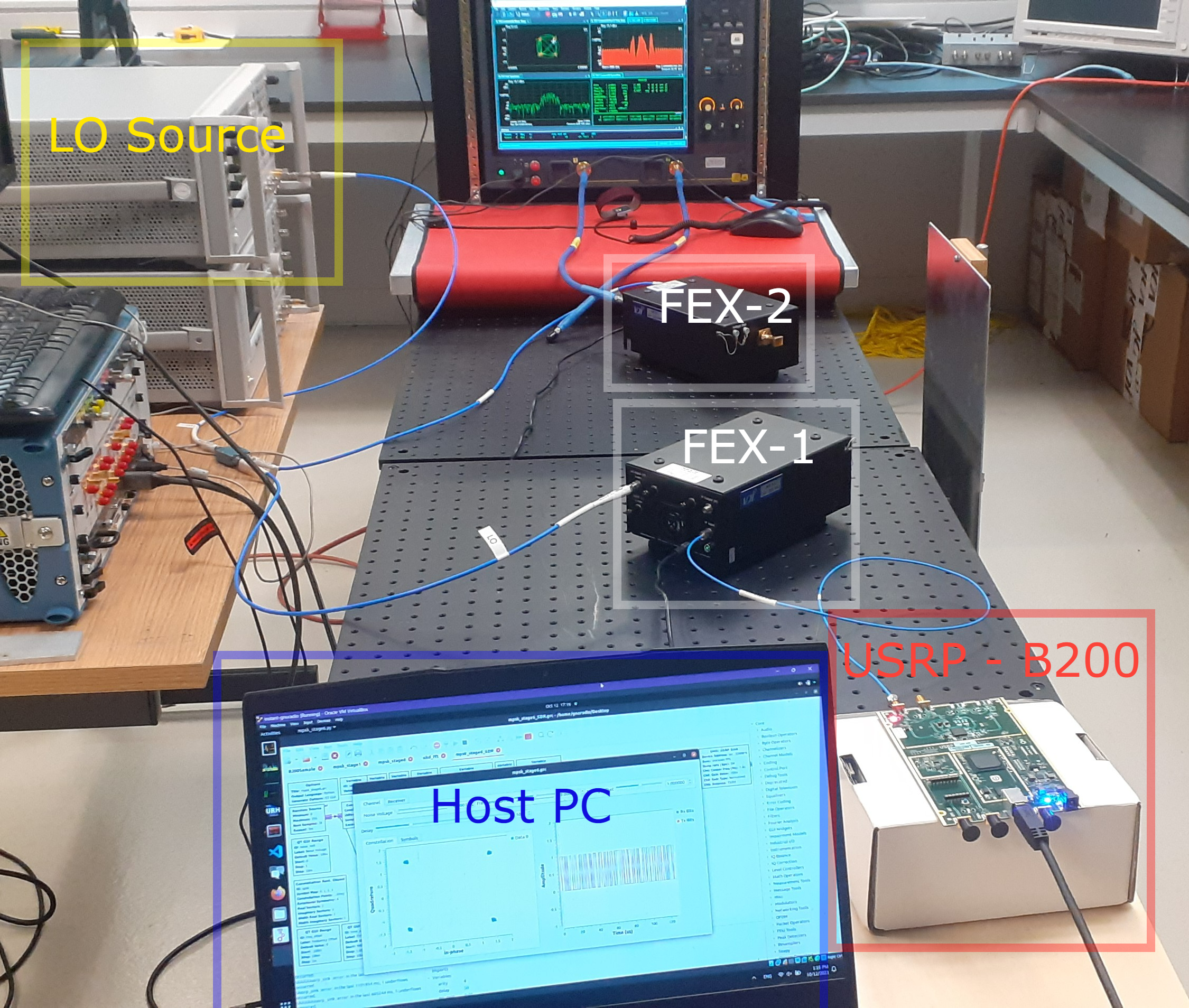}
    \caption{$180$ GHz SDR-THz testbed evaluating the non-line-of-sight extension with a reflector in PolyGrames.}
    \label{testbed}
        \vspace{-0.5cm}
\end{figure}
\begin{figure*}[] 
    \centering
    \includegraphics[width=0.8\textwidth]{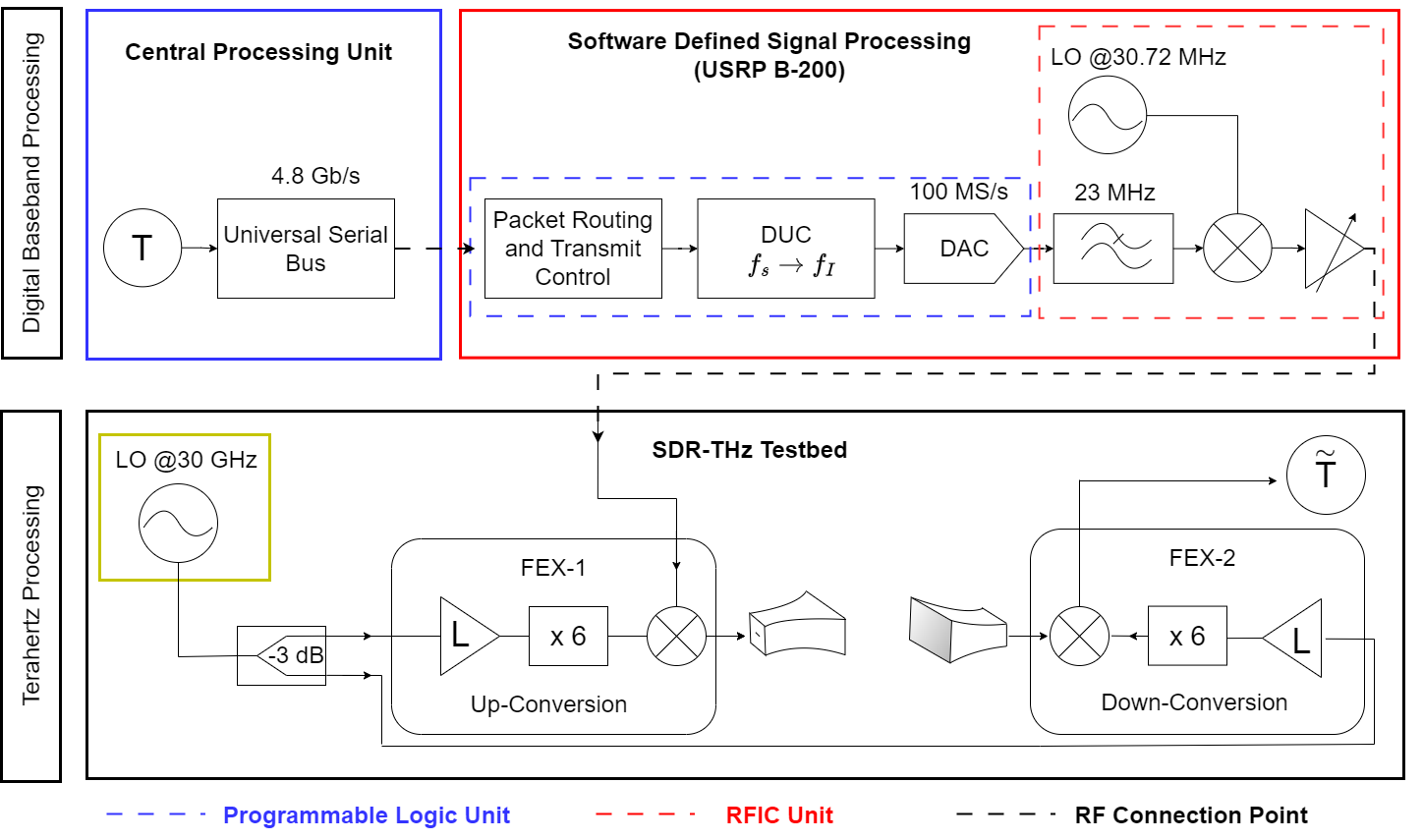}
    \caption{End to end SDR-THz testbed block diagram showing the several non-idealities on SDR output.}
    \label{blockdiag}
        \vspace{-0.5cm}
\end{figure*}

The cascaded stages of an SDR-THz testbed are shown in Figure \ref{blockdiag}. The baseband signal generation and the IF ($f_I$) up-conversion to sub-$6$ $\mathrm{GHz}$ are demonstrated by SDR transmission. Either filtered or unfiltered IF signal mixed with $M$ times multiplied local oscillator (LO) source on the frequency wave extension module (FEX-1). Consequently, a digitized over-the-air THz signal is generated\footnote{The double sidebands of $\pm f_I$ can be suppressed by a bandpass filter on FEX-1, yet this power loss has been disregarded for this study.} with the frequency $M\times f_\ell \pm f_I$ where $f_\ell$ is the LO frequency. On the reception, a secondary frequency wave extension module (FEX-2) is used for down-conversion of the received signal. End to end system model will be discussed deeper in Section \ref{exp}.

\subsection{A Cascaded SDR-THz System Model}
The generated signal on SDR meets with on-board hardware noise and an additional connection loss is introduced on the external radio frequency (RF) connection at the SDR output. The split LO source introduces $3$ $\mathrm{dB}$ insertion loss along with the phase noise. Lastly, the received signal is exposed to amplification on FEX-2. The stochastic model for the first stage ($S(1)$) output is the following: 
\begin{equation}
    S(1) \rightarrow \textbf{y}_{S_1}(t) = \sqrt{P_S} (\textbf{x}(t) + \textbf{w}^{tx}_I(t)) + \textbf{w}^{rx}_F(t),
\end{equation}
where $P_S$ is the SDR transmission power, $x \in \textbf{x}$ is the information symbol sequence with I/Q components as:
\begin{equation}
    x(t) = \mathrm{Re}\{[I(t) + j Q(t)] e^{j2 \pi f_c t}\}.
\end{equation}
On Eq. (1), $\textbf{w}^{tx}_I (t) \sim \mathbb{CN}(0,\kappa^2_I)$ is the SDR hardware noise on transmission and $\textbf{w}^{rx}_F(t) \sim \mathbb{CN}(0,P_S \kappa^2_{F,r})$ is the FEX-1 hardware noise on IF reception. The independent hardware impairment coefficients are denoted with $(\kappa_I, \kappa_{F,r} \geq 0) \in \mathbb{R}$. Simplifying the $S(1)$ as: 
\begin{equation}
    S(1) \rightarrow \textbf{y}_{S_1}(t) = \sqrt{P_S} \textbf{x}(t) + \textbf{n}_1(t). 
\end{equation}
Hereby, $\textbf{n}_s \sim \mathbb{CN}(0, P_S \kappa^2_S)$ where $\kappa_S = \sqrt{\kappa_I^2 + \kappa^2_{F,r}}$. Note that, if $\kappa_I = \kappa_{F,r} = 0$, then FEX-1 has only an amplified IF signal without noise. The signal-to-noise ratio (SNR) at the $S(1)$ becomes:
\begin{equation}
    \gamma_{(S1)} = \frac{1}{\kappa^2_S}.
\end{equation}
On the second stage ($S(2)$) from FEX-1 to FEX-2, the received signal model is the following:
\begin{equation}
    S(2)\rightarrow \textbf{y}_T(t) = \sqrt{\frac{P_T}{L}} \textbf{h}(t) \textbf{y}_{S_1}(t) +  \textbf{n}_2(t). 
\end{equation}
where $\textbf{n}_2(t) = \sqrt{\frac{P_T}{L}} \textbf{h}(t) \textbf{w}_F^{tx} (t) + \textbf{w}_{F2}^{rx} (t) + \textbf{z}(t)$, defining the $\textbf{w}_F^{tx}$ is the transmission noise on FEX-1, $\textbf{w}_{F2}^{rx}$ is the reception noise on FEX-2, and lastly, $\textbf{z} \sim \mathbb{CN}(0, N_0)$ is the additive complex white Gaussian noise with power spectral density of $N_0$. On this stage, $P_T$ is the FEX-1 transmission power, $L$ is the attenuation and conversion loss coefficient, $h \in \textbf{h}$ is the channel fading coefficient and, similar to $S(1)$, the aggregated noise is $\textbf{n}_2 \sim \mathbb{CN}(0,\frac{P_T}{L} h^2 \kappa^2_T + N_0)$ where $\kappa_T = \sqrt{\kappa^2_{F,t} + \kappa^2_{F2,r}}$.
SNR of $S(2)$ is the following:
\begin{equation}
    \gamma_{S(2)} = \frac{\frac{P_T |h|^2}{L \kappa^2_S N_0}}{\frac{P_T |h|^2 \kappa^2_S \kappa^2_T}{L N_0} +1}.
\end{equation}
Consequently, the SNR of the SDR-THz testbed is $\gamma = \gamma_{S(2)}$. On the LoS THz, poor multipaths on the channel result in a limited delay spread ($D$). Given the comparatively low sampling rate of SDR, the coherence bandwidth ($B \sim \frac{1}{D}$) exhibits remarkable width. This enables adherence to the flat fading assumption. Nevertheless, instantaneous channel estimation is not required, yet the statistical channel fading needs to be still estimated. 

\subsection{An $S(3)$ Extension with Reflector}
A portable reflector with a metal plate on the propagation way between FEX-1 and FEX-2 can significantly change the fading characteristics. In a sufficiently enough distance of reflector with $d_R$ the receiver does not allow the beam to diverge completely. As a result, orientation around FEX-1 and FEX-2 disrupts the LoS communication, leaving only the propagation path of the reflector. \textcolor{black}{Extending the the Equation (5) with diffuse scattering and material absorption on the reflector as}:
\begin{equation}
    L = (\xi + k a),
\end{equation}
where $\xi$ is the conversion loss, $k$ is the diffraction coefficient of the plate material and $a$ is the propagation loss of the path which is proportional to $d_R$ with a path loss exponent $\eta$ as $a \propto d_R^{-\eta}$. Furthermore, the LoS channel fading without a reflector can be obtained by surpassing the propagation time of each path. The channel coefficient with total of $K$ multipath with uniformly random phase assumption with $\phi \sim \mathbb{U}(0,2\pi)$ is the following:
\begin{equation}
h = \begin{cases}
    \sum_{i=1}^K k a_i \exp({-j2\pi f_c t_i}) \exp(j\phi), & \text{NLoS} \\
    \sum_{i=1}^K k a_i \exp(j\phi), & \text{LoS}   
\end{cases}
\end{equation}    
Note that LoS channel still contains multipaths without a reflector and $t_i = 0$ \cite{molisch2002capacity}. In addition, the distribution of the channel coefficients can be modeled with $\alpha-\mu$ distribution \cite{papasotiriou2021experimentally}. The probability distribution function is:
\begin{align}
    f(h) &= \frac{\alpha \mu^\mu \Bigl(\frac{h}{\beta}\Bigl)^{\alpha \mu -1} \exp{\Bigl(-\mu \Bigl(\frac{h}{\beta}\Bigl)^\alpha\Bigl)}}{\beta \Gamma\Bigl(\mu\Bigl)},
\end{align}
and the cumulative distribution function of $h$ is the following:
\begin{align}
    &F(h) = 1 - \frac{\Gamma\Bigl(\mu, \Bigl(\frac{h}{\beta}\Bigl)^\alpha \mu\Bigl)}{\Gamma\Bigl(\mu\Bigl)},
\end{align}
where $\alpha \in \mathbb{R}^+$ is the non-linearity in propagation environment, $\mu$ is the normalized variance of $h$ and $\beta = \sqrt[\alpha]{\mathbb{E}(h^a)}$.

\newsavebox\independent
\begin{lrbox}{\independent} 
\Large
    $\begin{aligned}
     P_f &= (\sum_{i=1}^{n}  P_i^{-1})^{-1} \\ 
      \hat{x}_f &= P_f (\sum_{i=1}^{n} P_i^{-1} \hat{x}_i)
    \end{aligned} $
\end{lrbox}
\newsavebox\correlated
\begin{lrbox}{\correlated}
   $ \begin{aligned} 
     P_f &= (e^T \Sigma^{-1} e)^{-1} \\
      \hat{x}_f &= P_f (e^T \Sigma^{-1} \hat{x})
    \end{aligned} $
\end{lrbox}
\newsavebox\skeww
\begin{lrbox}{\skeww}

   $ \begin{aligned}
     \frac{\pi}{2} - \arccos{\Bigl(\frac{I_M \cdot Q_M}{|I_M| |Q_M|}\Bigl)}
    \end{aligned} $
\end{lrbox}
\newsavebox\phas
\begin{lrbox}{\phas}

   $ \begin{aligned}
     \arctan{\Bigl(\frac{\Im \{I_0 + Q_0\}}{\Re \{I_0 + Q_0\}}\Bigl)} - \arctan{\Bigl(\frac{\Im \{I_M + Q_M\}}{\Re \{I_M + Q_M\}}\Bigl)} 
    \end{aligned} $
\end{lrbox}
\newsavebox\uc
\begin{lrbox}{\uc}

   $ \begin{aligned}
      \frac{\sqrt{\frac{1}{N} \sum_N (I_0 - I_M)^2 + (Q_0 - Q_M)^2 }}{\text{Norm Ref.}} \times 100\%
    \end{aligned}  $ 
\end{lrbox}
\newsavebox\ampdr
\begin{lrbox}{\ampdr}

   $ \begin{aligned}
      20\log{(E_{\text{av}})}, E_{\text{av}} = |E_2 \div E_1|
    \end{aligned}  $ 
\end{lrbox}
\newsavebox\freqq
\begin{lrbox}{\freqq}

   $ \begin{aligned}
      \text{abs}(f_c^0 - f_c^M)  
    \end{aligned}  $ 
\end{lrbox}
\newsavebox\gainnn
\begin{lrbox}{\gainnn}

   $ \begin{aligned}
      20 \log{\Bigl(\frac{|I_M|}{|Q_M|}\Bigl)}
    \end{aligned}  $ 
\end{lrbox}
\newsavebox\snrr
\begin{lrbox}{\snrr}

   $ \begin{aligned}
      10 \log{\Bigl(\frac{\sum_N (I_M + Q_M)^2 }{\sum_N (I_0 - I_M)^2 + (Q_0 - Q_M)^2 }\Bigl)}    
    \end{aligned}  $ 
\end{lrbox}
\begin{table}[]
  \caption{Error Metrics}
  \resizebox{1\columnwidth}{!}{%
  \begin{threeparttable}
\begin{tabular}{lp{6cm}@{\qquad}p{5cm}}
      Estimation Error Metrics & Estimation Rules \\ \midrule\midrule \\
        EVM [$\%\mathrm{RMS}$] & \usebox{\uc} \tnote{(*),(**)}  \\\\
     \cmidrule(l r){1-2}\\
    Phase Err. [$^\circ$] & \usebox{\phas} \tnote{(*)} \\\\ \cmidrule(l r){1-2}\\
    Skew Error [$^\circ$] & \usebox{\skeww} \tnote{(*)}  \\\\ \cmidrule(l r){1-2}\\
    Amplitude Droop [$\mathrm{dB/sym}$] & \usebox{\ampdr} \tnote{(*)} \\\\ \cmidrule(l r){1-2}\\
    Frequency Error [$\mathrm{Hz}$] & \usebox{\freqq} \tnote{(***)}  \\\\ \cmidrule(l r){1-2}\\
    IQ Gain Imbalance [$\mathrm{dB}$] & \usebox{\gainnn} \tnote{(*)}  \\ \\\cmidrule(l r){1-2}\\
    SNR [dB] & \usebox{\snrr} \tnote{(*)} \\\\ \midrule\midrule\\
    \end{tabular}

\begin{tablenotes}
  \item[(*)] $I_M \in [I_1, I_2,..., I_N]^T$ and $Q_M \in [Q_1, Q_2,..., Q_N]^T$ where $N$ is the length of symbols with symbol amplitude $E_i=\sqrt{(I_i + Q_i)^2}$. 
  \item[(**)] The peak magnitude of $I_0$ and $Q_0$ are used for normalization reference. 
  \item[(***)] Frequency Error encompasses carrier frequency offset, local oscillator clock offset, and sampling clock offset.
  \end{tablenotes}
  
  \end{threeparttable}
  }
  \label{tab}
  \vspace{-0.3cm}
  \end{table}

\section{Hardware Bottlenecks}
\label{sec3}
Two major challenges on THz are rate scarcity and frequency sparsity. To begin with, we define the issue of inadequate sampling in THz, along with potential hardware solutions. Subsequently, a concealed issue of frequency sparsity is brought to light through the definition of the realistic operating regions. \textcolor{black}{Relaxing these limitations not only reveals THz fading and inherent THz traits but also unlocks opportunities for multiplexing with very large antenna arrays and experimental evaluation of high data rates.}

\subsubsection{Rate Scarcity}
 Apart from the lack of THz supported clock frequency $f_\ell$ on legacy boards, onboard processing requires $f_s/f_c$ digital signal processing (DSP) to cover ultra-broadband. A THz operation with $B_T$ with the $\mathrm{GHz}$ order bandwidth to reach maximum data rate with $2 B_T \log{(M)}$ requires nearly Tbps data rate (e.g. $f_s = 800$ Gbps with $B_T = 100$ $\mathrm{GHz}$ for $M=16$) DSP on both transceiver and receiver. Taking the best case scenario of $f_\ell$ with $8-10$ $\mathrm{GHz}$ application specific integrated circuit (ASIC), the conventional signal processing method fails and creates a bottleneck for such broadband transmission. Parallel pipeline structures and artificial intelligence can offer a solution to this conundrum. Similarly, a host supported SDR can be used for hybrid real-time \& post-processing without overflowing the data stream for various use cases. A second constraint, the discrepancy in broadband filters presents a major drawback. Roll off characteristics, phase distortions and nonlinear effects of filter components reduce the signal quality to an unpredictable degree.

\subsubsection{Frequency Sparsity}
One major challenge for ultra-broadband transmission over the THz region is the constrained windows for finite regions of operation. Due to the critical absorption and scattering, THz windows limit the effective use of such broadband. This phenomenon enhances the significance of optimizing efficient utilization within the THz spectrum. However, this is not the only band limit on THz.  

One method for reaching THz is extending the operating frequency via multiplication. Frequency sparsity poses a practical challenge when operating in the THz range at a non-ideal center frequency. Simply, an operating frequency region $g$ can be defined with linear function as follows: 
\begin{equation}
    g = f_{Ex} + \Delta f_{Ex}, \forall f_{Ex}\in \mathbb{Z^+},
\end{equation}
where $f_{Ex}$ is the nominal frequency term and $\Delta f_{Ex}$ is the smallest frequency step. Recalling the extended frequency $M\times f_\ell \pm f_I$, defining the step size for each term as $\Delta M \in \mathbb{N}$, $\Delta f_\ell \in \mathbb{R}$ and $\Delta f_I \in \mathbb{R}$. A frequency extender with $\Delta M >> \Delta f_I$ and $\Delta M >> \Delta f_\ell$ can only reach the following operation region:
\begin{align}
    g &= (M + \Delta M) \times (f_\ell + \Delta f_\ell) + (f_I \pm \Delta f_I).
\end{align}
Cancelling the nominal terms of $f_{Ex}$ results the $\Delta f_{Ex}$: 
\begin{equation}
    \Delta f_{Ex} \approx M \Delta f_\ell + \Delta M \Delta f_\ell \pm \Delta f_I.  
\end{equation}
Equation (13) gives the approximation for frequency resolution of $g$. Repeating the identical steps for an ideal bandpass filtered $g^p$ case with image rejection is the following:
\begin{align}
  \Delta f^p_{Ex} \approx M \Delta f_\ell + \Delta M \Delta f_\ell.  
\end{align}
Without loss of generality, introducing a broadband constraint  $\epsilon \in \mathbb{R^+}$ with a linear relationship between $\Delta f_\ell$ and $f_\ell$ such as $\epsilon = \frac{\Delta f_\ell}{f_\ell}$. 
Thus, a generic frequency step definition for $g^p$ is given below:
\begin{equation}
    \Delta f^p_{Ex}\approx \begin{cases}
			0, & \text{$\epsilon = 0$, }\\
            \Delta f_\ell(M + \Delta M), & \text{otherwise.}
		 \end{cases}
\end{equation}
As a result, the selection of operating region is expected to have at least $\Delta f_{Ex}^p$ step size.   

\section{Experimental Results}
\label{exp}

A real-time SDR-THz performance evaluation has been conducted in a lab environment. Two stages of single input single output (SISO) demonstrated with one Universal Serial Radio Peripheral (USRP) B200 as a transmitter has been used in a single RF channel. USRP has a programmable logic unit is Spartan-6 XC6SLX75 and it utilizes an RFIC unit of Analog Devices AD9364. Two identical N9029AV05 model frequency extenders for FEX-1 and FEX-2 have been used for up and down conversion respectively. One E8267D pulse signal generator used as an LO source for both extenders. A UXR0502A signal analyzer at the FEX-2 end collects the data stream. The external physical connections were done by $2$-meter coaxial cables. FEX-1 and FEX-2 shares a co-shares an LO source via a wideband ($6-40$ $\mathrm{GHz}$) two-way splitter operating in $f_\ell = 30$ $\mathrm{GHz}$ and the mixed signal up-converted to $180 \pm 2.4$ $\mathrm{GHz}$ and transmitted with $P_T = 5$ $\mathrm{dBm}$. Both FEX-1 and FEX-2 are equipped with directive horn antennas having $11$ $\mathrm{dBi}$ antenna gain, $3$ $\mathrm{dB}$ beamwidth with $8.9^\circ$ on E-plane and $10.28^\circ$ on H-plane. The aperture efficiency of identical antennas is notated as $0.511$. A power meter measured the total insertion loss as $11$ $\mathrm{dB}$. Over the air signal captured by the FEX-2 to be down-converted to IF. The down-converted IF signal is synchronized and analyzed over the signal analyzer.

  \begin{figure}[t] 
    \centering
    \includegraphics[width=1\columnwidth]{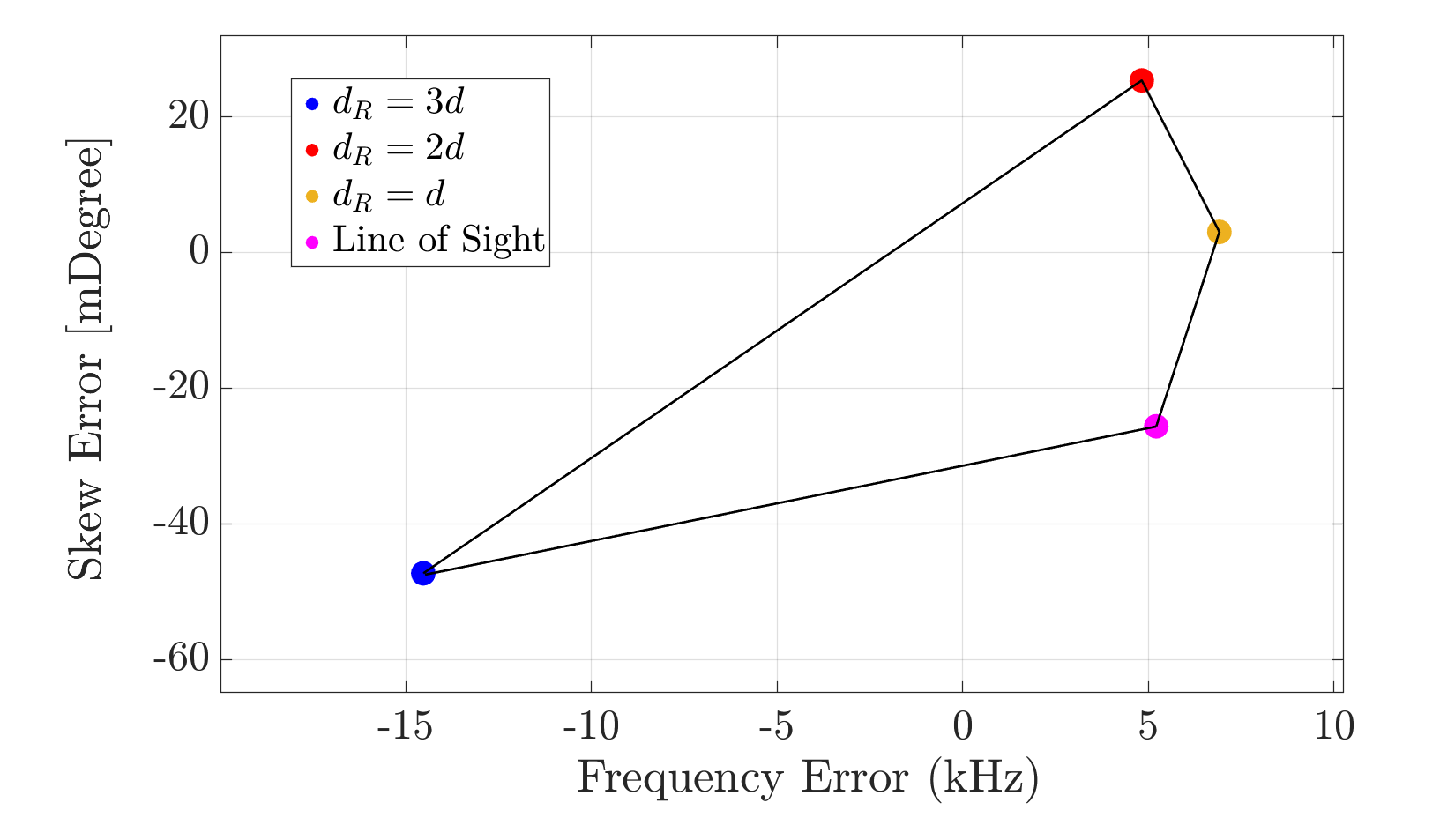}
        \vspace{-0.5cm}
    \caption{Skew error and frequency error for each measurement.}
    \label{skew}
    \vspace{-0.5cm}
\end{figure}

\subsection{Measurement Details}
The tests were carried out in $4$ distinct segments. In order to see the impact of $d_R$, the reflector is kept orthogonal to LoS path with the distances of $d_R \in \{d, 2d, 3d\}$. First $3$ tests utilize the SDR-THz testbed with reflector extension for each $d_R$ where $d=12.7$ $\mathrm{cm}$, $2d=17.7$ $\mathrm{cm}$ and $3d=22.7$ $\mathrm{cm}$. On the last test, LoS measurement was conducted without a reflector with a distance of $d=15.24$ $\mathrm{cm}$. The clock rate of the SDR is $30.72$ MHz and pseudo-random $10.000$ bits are generated repeatedly. SDR is operated with $f_s = 2$ $\mathrm{MS/s}$ and quadrature phase shift keying (QPSK) modulation is performed. A root raised cosine filter is used for shaping the rectangular pulse train. Thereby, the SDR output corresponds to roughly $0.8$ $\mathrm{MHz}$ bandwidth. The IF signal on $2.4$ $\mathrm{GHz}$ amplified with $P_S = 5$ $\mathrm{dBm}$ and directed towards FEX-1 to be up converted to THz with $M=6$ mixer multiplication. Continuing from this, the demodulation process involves the signal recovery steps, but the individual bits are not analyzed. 

\subsection{Performance Evaluation}

\begin{figure}[t] 
    \centering
    \includegraphics[width=1\columnwidth]{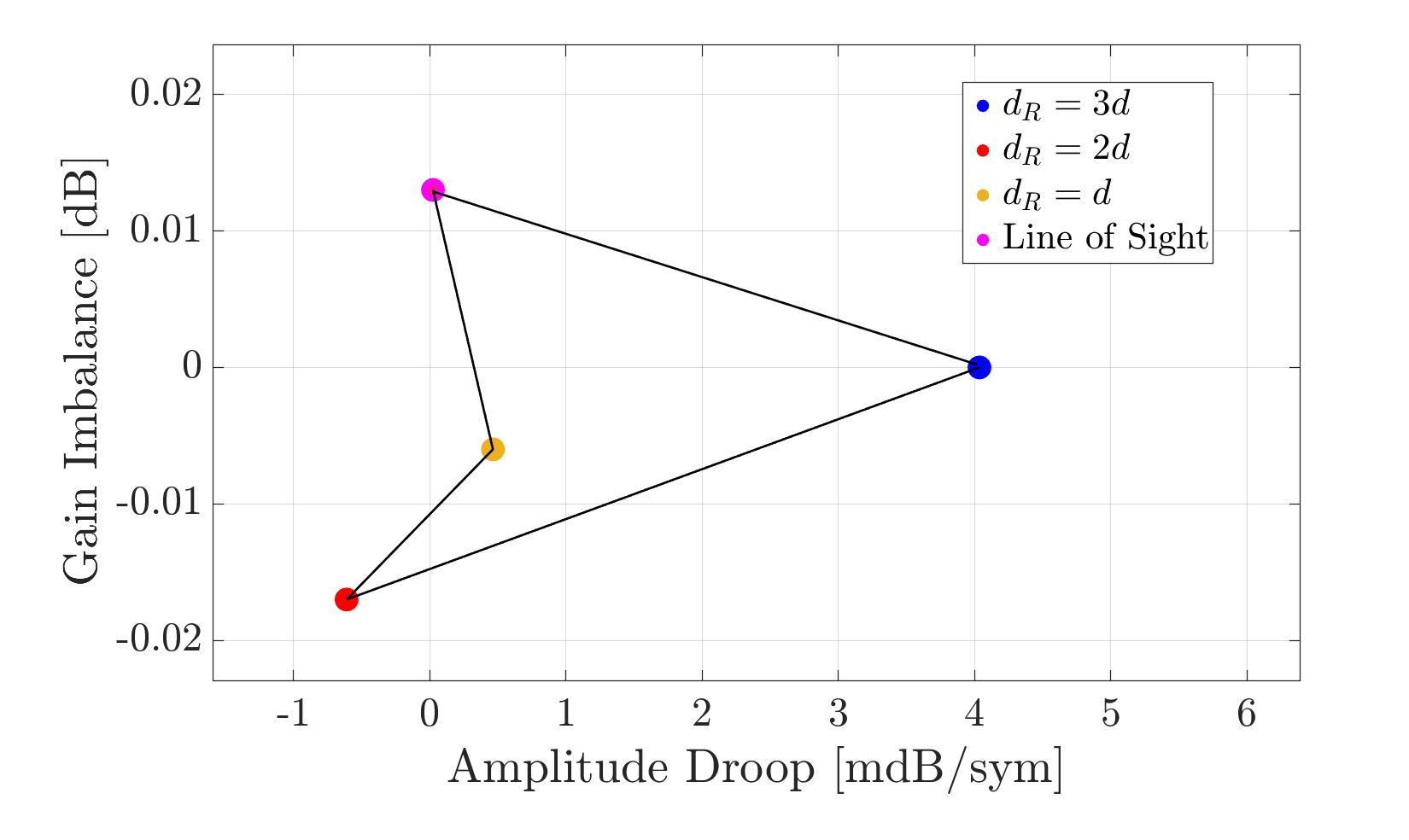}
        \vspace{-0.6cm}
    \caption{Amplitude droop and gain imbalance for each measurement.}
    \label{amp}
        \vspace{-0.5cm}
\end{figure}

\begin{figure}[] 
    \centering
    \includegraphics[width=0.5\textwidth]{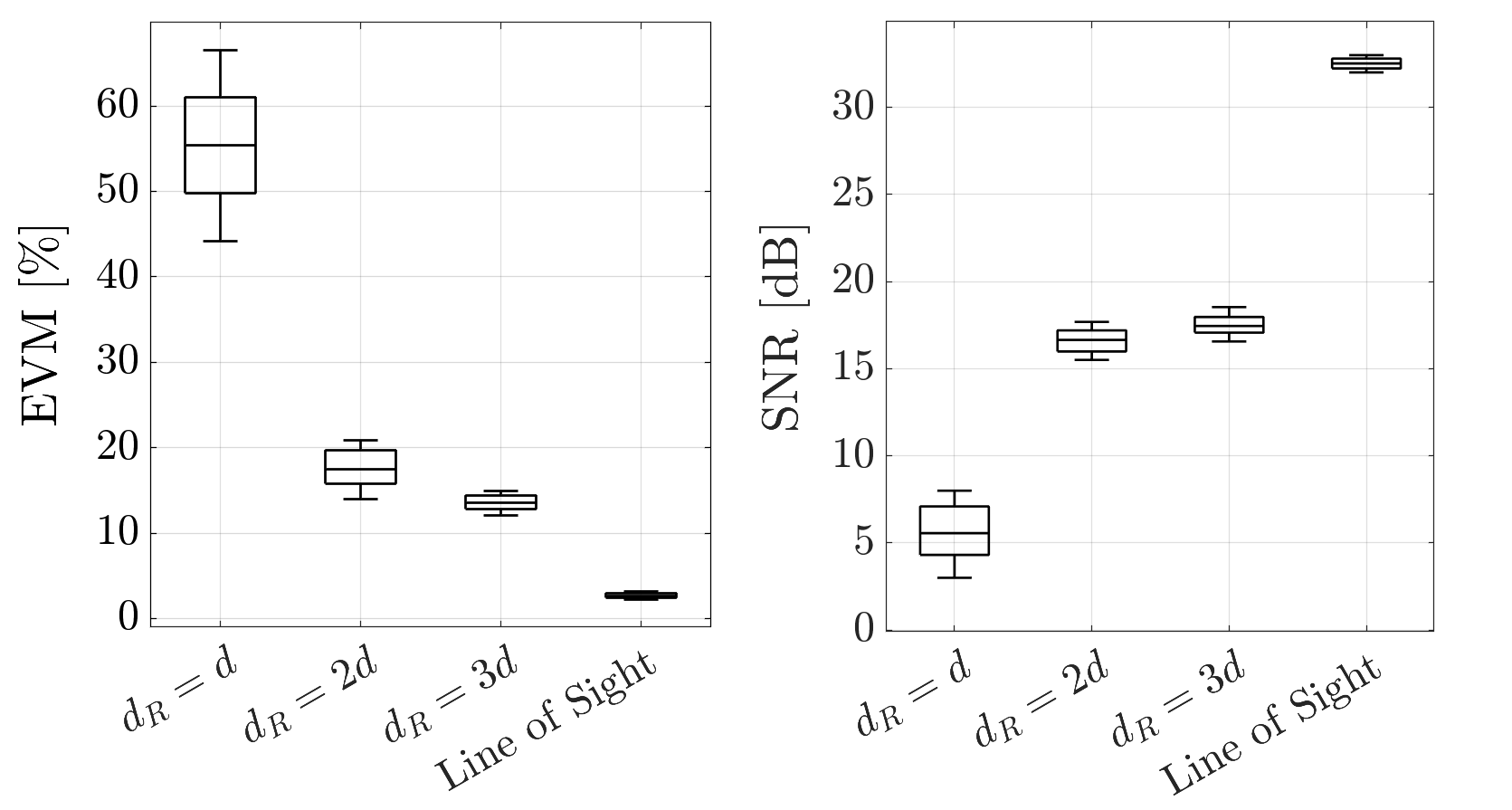}
        \vspace{-0.5cm}
    \caption{A boxplot for EVM and SNR measurements for each measurement.}
    \label{box}
    \vspace{-0.5cm}
\end{figure}

The designed SDR-THz system model results are given in Figure \ref{blockdiag}. Correspondingly, the error metrics obtained during the measurement are given in the Table \ref{tab}. Root-mean-square averaging was performed with $N=1000$ to approximate the error metrics. Figure \ref{box} shows that reflector distance has a negative relationship with the signal quality. The absence of surface roughness serves as compelling evidence that the loss of the reflector is primarily attributed to material loss and increased path loss. 

Figure \ref{skew} depicts that the frequency error on SDR-THz is in the order of $\mathrm{kHz}$ and saturated for $d_R=\{d,2d\}$, whereas the diffuse scattering at $d_R = 3d$ predominates the specular ray and leading to dispersive channels for Gaussian beams. On the other hand, the skew error on a constellation is stable with less than $1^{\circ}$ for all cases. Interestingly, the reflector can mitigate the skew and frequency error for $d_R = (d,2d)$ at the expense of considerable SNR losses. Nevertheless, it has been observed that this trade-off becomes less advantageous as the reflector distance increases to $d_R = (3d)$.  

\begin{figure}[] 
    \centering
    \includegraphics[width=1\columnwidth]{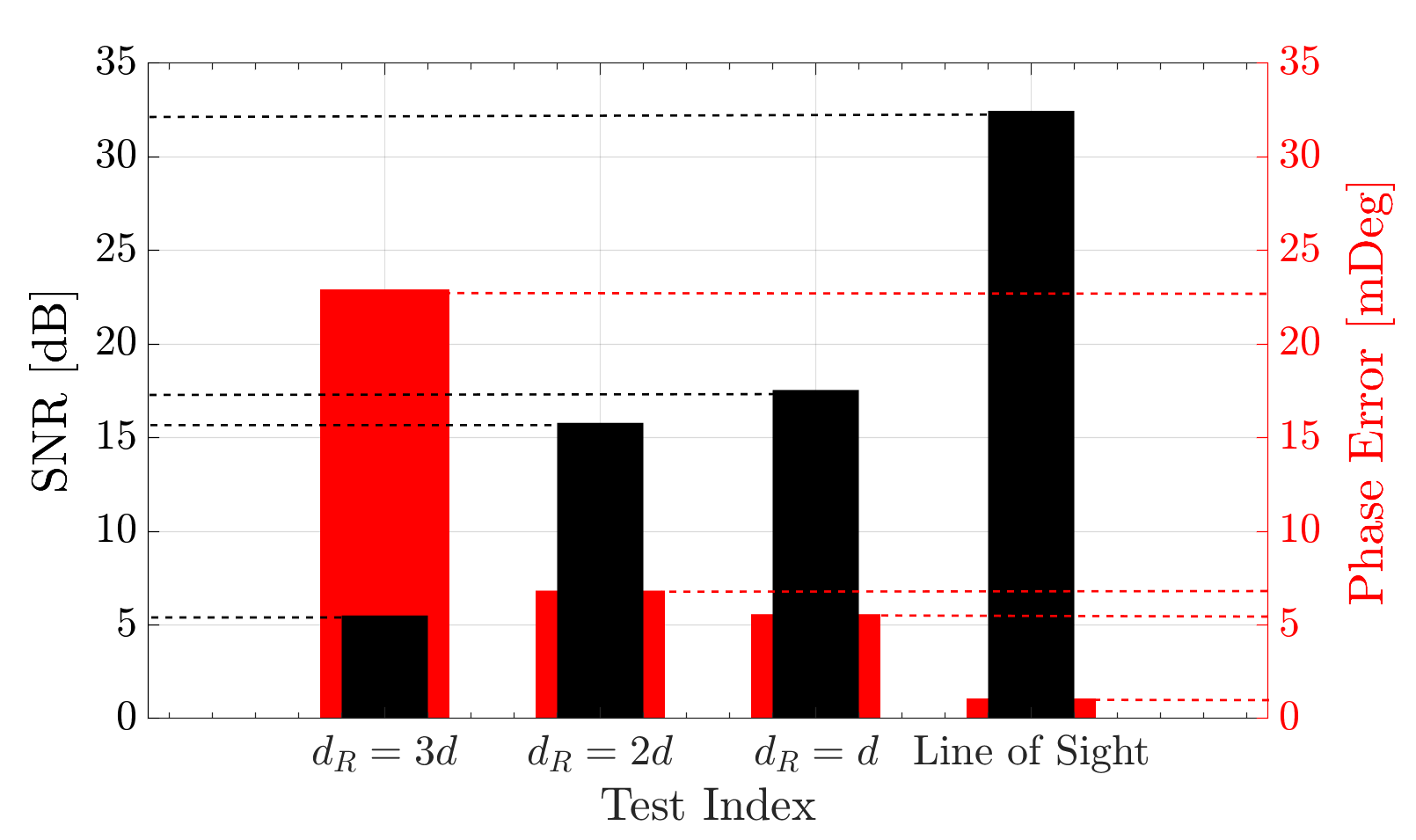}
        \vspace{-0.5cm}
    \caption{SNR and phase error correlations for each measurement.}
    \label{evm}
        \vspace{-0.5cm}
\end{figure}

As illustrated in Figure \ref{evm}, the SNR, EVM, and phase error exhibit an anticipated pattern corresponding to the $d_R$ e.g the LoS experiment has an average SNR of $32.463$ $\mathrm{dB}$ while the use of a reflector lowers it down to $17.55$ $\mathrm{dB}$ and below. Nevertheless, the amplitude droop for the first captured symbol energy ratios is stable and close to ideal and around $0$ 
$\mathrm{dB}$, while the amplitude droop is seen on the long distance reflector increases dramatically up to $4$ $\mathrm{mdB/sym}$. 

We point out the importance of not using external sources for clock synchronization, mitigating the LO sourced timing offsets. Note that, the channel is non-disperse due to the single path and stationary nodes meeting the previously mentioned flat fading channel assumption for LoS test. The signals do not possess pilot symbols, therefore a symbol timing estimation is not required. Nevertheless, a blind channel estimator is used to estimate channel frequency response, followed by a zero forcing channel equalization. This marks the completion of the signal processing at the reception, which is subsequently followed by the calculation of error metrics.
\label{sec4}

\section{Conclusion}        
\label{sec6}
This study demonstrates a real-time SDR-THz testbed and evaluates the system performance with I/Q error metrics. First, a generic SDR-THz system model has been derived and practical THz hardware bottlenecks are investigated. Then, the instrumentation and reflector aided measurement results are discussed. It has been observed that  $180$ $\mathrm{GHz}$ radiation with $3.2$ $\mathrm{Mbps}$ can reach up to $32.463$ $\mathrm{dB}$ with less than $0.8$ $\mathrm{mdeg}$ phase error and $<1^{\circ}$ skew error in line-of-sight. On the other hand, the use of a reflector in signal quality enhancement is bottlenecked in the tests for $d_R = 3d$.     

\textcolor{black}{
The realm of SDR-THz harbors numerous unexplored avenues, waiting to be uncovered. Real-time experiments in secure THz communication, multi-input-multi-output (MIMO) communication, diverse multi-access methods, and THz spectroscopy stand out as untapped domains. However, it is the ultra-broadband SDR-THz that presents the most critical area of study, encompassing significant challenges as discussed in Section \ref{intro}. For instance, the security dimension of THz communication can be more thoroughly explored with a larger spectral resource, enabling the implementation of diverse coding schemes and more intricate key generation techniques. Moreover, SDR-THz holds promise as a tool for cognitive radio applications, opening doors to innovative possibilities. In a similar vein, the fusion of sensing and communication within the THz spectrum offers several exciting prospects, such as localization and mapping, waiting to be explored.
}
\section{Acknowledgement}
This work was supported in part by John R. Evans Leader Fund (JELF) of Canada Foundation of Innovation (CFI). 
\bibliographystyle{IEEEtran}
\bibliography{reference}

\end{document}